\documentclass[acmsmall]{acmart}

\usepackage{hyperref}
\usepackage{tabularx}
\usepackage[shortlabels]{enumitem}

\AtBeginDocument{%
  \providecommand\BibTeX{{%
    \normalfont B\kern-0.5em{\scshape i\kern-0.25em b}\kern-0.8em\TeX}}}

\def\toolname{\textsc{SUMMIT}}

\newcommand*\circled[1]{\raisebox{.5pt}{\textcircled{\raisebox{-.9pt}{#1}}}}

\begin{document}

\setcopyright{none}
\acmJournal{PACMHCI}
\acmYear{2023} \acmVolume{7} \acmNumber{CSCW2} \acmArticle{297}
\acmMonth{10} \acmPrice{15.00}\acmDOI{10.1145/3610088}

\received{July 2022}
\received[revised]{January 2023}
\received[accepted]{March 2023}

\title[SUMMIT: Scaffolding OSS Issue Discussion Through Summarization]{\toolname{}: Scaffolding Open Source Software Issue Discussion Through Summarization}

\author{Saskia Gilmer}
\affiliation{%
 \institution{McGill University}
 \city{Montreal}
 \state{Quebec}
 \country{Canada}}
\email{saskia.gilmer@mail.mcgill.ca}

\author{Avinash Bhat}
\affiliation{%
 \institution{McGill University}
 \city{Montreal}
 \state{Quebec}
 \country{Canada}}
\email{avinash.bhat@mail.mcgill.ca}

\author{Shuvam Shah}
\email{shuvam.shah@polymtl.ca}
\affiliation{%
 \institution{Polytechnique Montreal}
 \city{Montreal}
 \state{Quebec}
 \country{Canada}}
 
\author{Kevin Cherry}
\affiliation{%
 \institution{McGill University}
 \city{Montreal}
 \state{Quebec}
 \country{Canada}}
\email{kevin.cherry@mail.mcgill.ca}

\author{Jinghui Cheng}
\email{jinghui.cheng@polymtl.ca}
\affiliation{%
 \institution{Polytechnique Montreal}
 \city{Montreal}
 \state{Quebec}
 \country{Canada}}

\author{Jin L.C. Guo}
\email{jguo@cs.mcgill.ca}
\affiliation{%
 \institution{McGill University}
 \city{Montreal}
 \state{Quebec}
 \country{Canada}}

\begin{abstract}
Issue Tracking Systems (ITS) often support commenting on software issues, which creates a space for discussions centered around bug fixes and improvements to the software. For Open Source Software (OSS) projects, issue discussions serve as a crucial collaboration mechanism for diverse stakeholders. However, these discussions can become lengthy and entangled, making it hard to find relevant information and make further contributions. In this work, we study the use of summarization to aid users in collaboratively making sense of OSS issue discussion threads. Through an empirical investigation, we reveal a complex picture of how summarization is used by issue users in practice as a strategy to help develop and manage their discussions. Grounded on the different objectives served by the summaries and the outcome of our formative study with OSS stakeholders, we identified a set of guidelines to inform the design of collaborative summarization tools for OSS issue discussions. We then developed \toolname{}, a tool that allows issue users to collectively construct summaries of different types of information discussed, as well as a set of comments representing continuous conversations within the thread. To alleviate the manual effort involved, \toolname{} uses techniques that automatically detect information types and summarize texts to facilitate the generation of these summaries. A lab user study indicates that, as the users of \toolname{}, OSS stakeholders adopted different strategies to acquire information on issue threads. Furthermore, different features of \toolname{} effectively lowered the perceived difficulty of locating information from issue threads and enabled the users to prioritize their effort. Overall, our findings demonstrated the potential of \toolname{}, and the corresponding design guidelines, in supporting users to acquire information from lengthy discussions in ITSs. Our work sheds light on key design considerations and features when exploring crowd-based and machine-learning-enabled instruments for asynchronous collaboration on complex tasks such as OSS development.
\end{abstract}



\maketitle

\section{Introduction}
Issue tracking systems (ITSs) play a central role in open source software (OSS) maintenance and evolution, supporting a variety of use cases for different stakeholders. End-users often rely on issues to report problems they encounter when using the software and to voice their opinions on future improvements. Developers use issue discussions to spot shortcomings and troubleshoot with other contributors; they also take part in these discussions to better understand end-users' needs. Project owners and maintainers use these threads to track progress and communicate recent development updates to a broader audience. The resulting issue discussion threads in the ITSs become a living project documentation that records rich information about the collaborative progress of the OSS from diverse perspectives~\cite{aghajani2020Software, Arya2019}.

Existing ITSs, such as GitHub Issues~\cite{github_issue}, display comments on issue discussion threads chronologically. Users can go through the threads via a vertical scroll bar when the content cannot fit the screen. As OSS grows in popularity, issue discussion threads can contain hundreds of comments, making it hard to find relevant information and make new contributions on top of the existing discussions. Even with shorter issue threads, key points may be hidden in the regular comment text and difficult to discern. Our initial investigation of issue threads from popular OSS repositories reveals that such a problem intensifies as more users join the discussion with different backgrounds, perspectives, and levels of familiarity with the topic. Commenters employ various strategies to facilitate information retrieval for the readers, including using large fonts to increase the visibility of important information, locking the thread to limit its length, and summarizing comments. Yet the sophistication of these strategies is limited by the user interface of current ITSs. 

\begin{figure} [t!]
    \centering
    \includegraphics[width=0.95\textwidth]{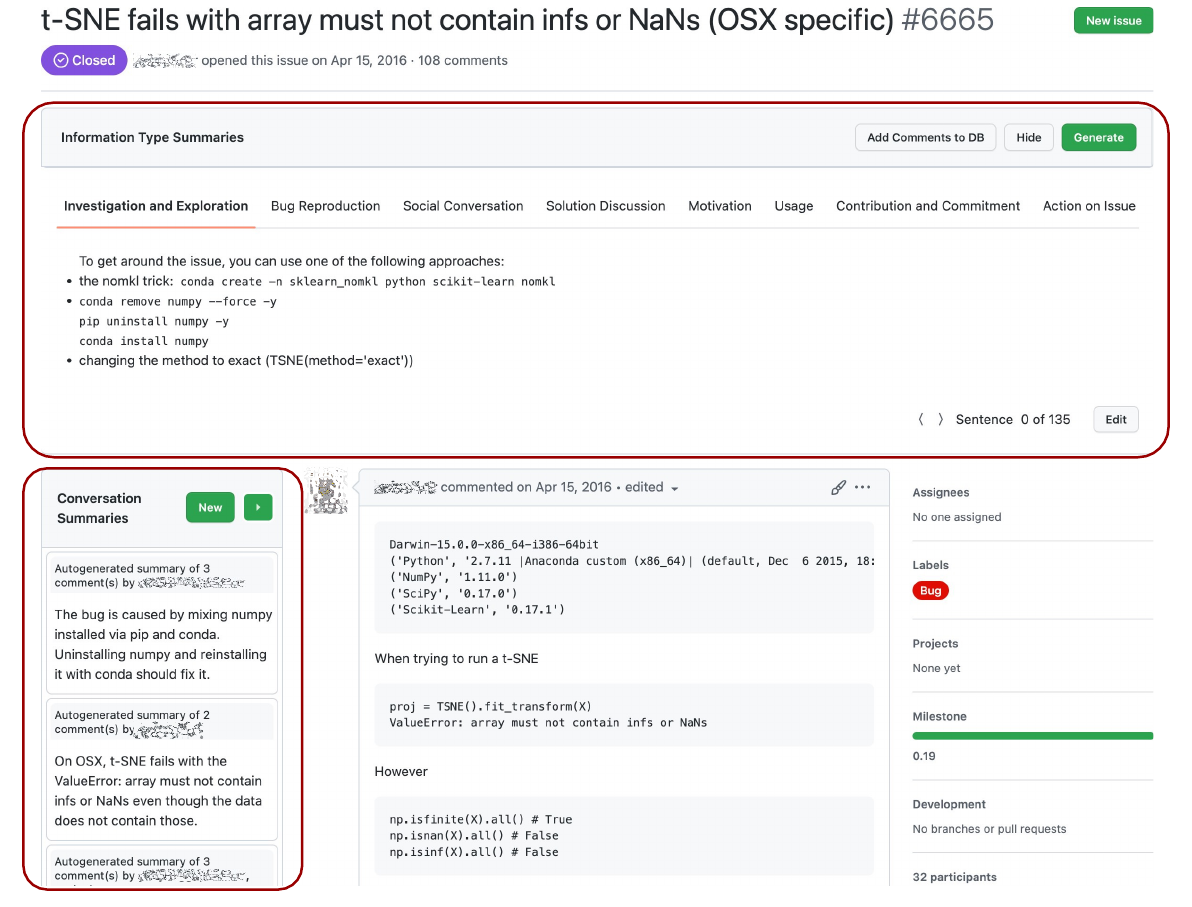}
    \caption{Two panels are added to the issue interface when users access an issue on GitHub (indicated by the red boxes), i.e. Information Type Summaries (on top) and Conversation Summaries (as the left panel).}
    \label{fig:summit_overview}
\end{figure}

Meanwhile, previous work has discussed the potential benefits of adopting summarization or content recommendations to help users get oriented in online discussions. Summarizing discussion threads lowers the barrier to commenting and therefore encourages users to contribute their ideas~\cite{10.1145/3449246}. Such sharing of ideas has been shown to boost the level of innovation and quality of the final work~\cite{zhang2017wikum}. Highlighting opposite views on the same issues can further increase people's awareness of different stances and opinions~\cite{10.1145/3172944.3172970, wang2020argulens}. 
Previous efforts, however, often focus on forums of particular topics such as education, politics and science. There lacks a clear understanding and consensus on the kind of information worth summarizing for \textit{software issues}, that are centered on user feedback and engineering progress management. It is also unclear how to support the ITS users to keep up with the fast-evolving discussions through summaries while making the manual effort manageable, and at the same time, fit the summarization practice into the existing workflow of OSS stakeholders.
 
In this work, we investigate these problems. We particularly focus on exploring ways to scaffold software issue discussions through collaborative summarization methods. The term ``scaffold'' refers to any assistance provided by the user interface to support the OSS stakeholders to access and create content in the thread and, therefore, foster their discussion and collaboration.
Toward this end, we formulate three concrete research questions (RQs), each of which is built on the investigation of the previous one. 
We started by asking: \textbf{What are the characteristics of existing summaries that naturally occurred in the OSS issue discussion threads (RQ1)?}
By analyzing long threads from popular OSS projects to understand the existing practice of summarization, we found that users of GitHub Issues already devote a significant effort to summarizing comments. In the 30 discussion threads examined, we found 108 summaries. However, the manual effort of creating those summaries was daunting; one user even mentioned that they spent three hours writing their summarizing comment. Our content analysis resulted in three dimensions to characterize the summaries from issue threads, i.e., summary objective, style, and target audience. These dimensions serve as the basis to envision the use cases of summaries in ITSs. 

A clear depiction of current summarization practice and the role of summaries served in the discussion threads led us to ask: \textbf{What kind of support can be provided to satisfy the need to write and use summaries in OSS ITSs (RQ2)?} Combining the empirical investigation of existing summaries and previous work on ITS and online discussion forums, we proposed a conceptual design of \toolname{}, a tool that allows collaborative writing, editing, and retrieval of summaries. Instrumented by an early prototype of \toolname{}, we conducted a formative user study with different OSS stakeholders and distilled a set of guidelines to inform the design of tools that leverage collaborative summarization to support authoring summaries and acquiring information in OSS issue discussions.

We then designed and implemented \toolname{} as a browser plugin following the design guidelines. \toolname{} adds two summary panels on the issue thread, i.e., \textit{Information Type Summaries} and \textit{Conversation Summaries} (see Fig.~\ref{fig:summit_overview}). The two types of summaries enable the users to glance through and retrieve information about various aspects of the issue discussions with different levels of granularity. Our last research question, therefore, is about understanding: \textbf{How does \toolname{} affect OSS stakeholders' behavior and perception of the information acquisition process (RQ3)?} Through a user study with 16 participants, we evaluated \toolname{} from the perspectives of OSS end-users and contributors. Compared with the existing ITS interface design, \toolname{} brings a clear advantage in \textit{reducing the mental effort for ITS users and restructuring the information to direct users with different information needs}.

Our work makes the following contributions: (1) we provide empirical evidence on how summarization has been used to facilitate software issue discussions; (2) we suggest a set of guidelines for designing tools for collaborative authoring and using summaries in ITSs; (3) we propose a novel user interaction design of ITS anchored on two types of summaries for OSS issue discussions and use natural language processing techniques to alleviate the extensive manual effort required for managing the summaries; and (4) our findings from the user study highlight the opportunity to explore crowd-based and machine-learning-enabled instruments for asynchronous collaboration on complex tasks such as building and maintaining open source software.

\section{Background and Related Work}

In this section, we discuss how our work is contextualized in the previous effort focused on investigating (1) the role of issue discussion threads in OSS development, (2) online collaboration platforms and tools, and (3) automated summarization techniques.

\subsection{Issue Discussion Threads for Software Projects}
``Software issue'' is an umbrella term that can refer to bug reports, feature requests, improvement suggestions, or any other topics raised by software stakeholders~\cite{nicholson2020traceability}. Due to the evolving nature of software development and maintenance, issues are increasingly important to host discussions and build communities among different software stakeholders, in particular for open source software projects that heavily rely on asynchronous remote collaboration~\cite{Arya2019}. Issues are usually managed in issue tracking systems (ITSs), such as GitHub Issues~\cite{github_issue}. After being posted on the ITSs, the discussions surrounding the original issues often develop continuously and are referred to in other contexts (such as in other issue threads or code contributions)~\cite{nicholson2021issue}. 

Previous work on ITSs focuses on problems such as understanding the types of issues and their impact on the project~\cite{6698918}, identifying the relationship between issues~\cite{nicholson2021issue}, and recommending issues to relevant maintainers~\cite{10.1145/3510003.3510196}. Recent attention is also devoted to understanding the rich discussion generated surrounding individual issues. For example, \citet{Arya2019} identifies 16 types of information contained in issue threads that might serve different stakeholders for various tasks. Manually discovering relevant information types in the thread, however, is challenging, especially when the topic is complex and the thread is lengthy. While \citet{Arya2019} demonstrated the potential of using machine learning methods to classify the information types of each sentence in the thread, no tool exists that allows users to interact with the information types directly. Their classification techniques have been adapted to dissect the discussion by extracting the argumentation structure from the thread~\cite{wang2020argulens}, but the discussion is limited to the usability aspect of the software. 

Our work builds on the previous literature to support the information discovery in issue threads saved in ITSs. Particularly, our \toolname{} tool leverages two different summarization mechanisms to enhance the information organization of issue threads and facilitate information acquisition for ITS stakeholders. Parts of \toolname{} directly extend the work by \citet{Arya2019}, allowing the users to effectively retrieve the most relevant information for their tasks at hand.

\subsection{Online Collaboration Support}
The designs of existing ITSs normally incorporate rich metadata to record important information related to the issue; examples of such metadata include labels indicating the properties of the issue, responsible programmers to whom the issue is assigned, the corresponding project milestones, etc. Nevertheless, the main body of the issue discussion still resembles the design of many online forums, in which the comments are organized chronically. It consequently suffers from many of the same problems as online forums, such as miscommunication~\cite{storey2016social}, negative feedback loops~\cite{cheng2014community}, and lack of inclusiveness~\cite{igor2013newcomers}. In the CSCW community, various means have been proposed to address those problems, normally separately, for different domains. For example, to combat misinformation related to health, \citet{10.1145/3274327} built a classifier to identify Twitter users who might propagate misinformation using features about the user profile and their tweets. To mitigate the echo chamber effects, \citet{10.1145/2531602.2531711} suggested the potential to show the stands and magnitudes of positions on controversial topics at different stages of users' information consumption in online discussions. 

Related to our research, \citet{zhang2017wikum} proposed a recursive summarization tool to support online collaboration in general forums. The main idea is to combine the design of Wikipedia and forums so that the discussions can build on previous summaries contributed by individual users. The summaries are organized in trees to incorporate different levels of discussion. In their user study, both the editors and readers found the proposed tool productive to create and digest summaries. However, they also observed the tension between the summarization and navigation goals with the complicated tree structure. Our tool design is partially inspired by \citet{zhang2017wikum}'s work. However, our objective is contextualized in OSS communities, i.e., to enhance the communication on issue threads while assisting well-established use cases of ITSs. Therefore, we investigate the design to support concrete information seeking needs that fit into the existing issue discussion workflows.

\subsection{Automated Summarization}
Automated summarization techniques have been extensively explored by the research on natural language processing to alleviate the strenuous effort of manual summarization. The techniques can be broadly categorized into extractive and abstractive approaches; extractive approaches rank and select the most important sentences from the input text~\cite{nallapati2017summarunner} and abstractive approaches synthesize summaries from the input text without direct copying~\cite{lin2019abstractive}.  Both categories of methods have been applied to applications such as summarizing online opinions towards news and product reviews~\cite{gerani2014abstractive}, medical notes and reports~\cite{zhang2020Optimizing, krishna2021Generating}, and government reports~\cite{huang-etal-2021-efficient}, among others. 

Summarization of software issues poses several unique challenges to the existing methods. First of all, since a large number of information types exist in the issue thread and can serve different stakeholders at various stages of the issue lifespan, it is difficult to generalize the existing methods to new issues or even the same issue at a different stage. Moreover, the quality of the summaries needs to be controlled to maintain the integrity of the discussion and the efficiency of the information reuse. Considering those aspects, our work uses an extractive summarization model to link the output summary directly to the sentences in the original input text and therefore design functions that support the inspection of the summary quality and provenance.  In particular, we adopt a summarization method called BERTSumExt as the backbone of our tool~\cite{liu-lapata-2019-text}. BERTSumExt is a state-of-the-art machine-learning based method that builds on the pre-trained language model BERT~\cite{devlin-etal-2019-bert}. At the same time, our tool is general enough to adopt more recent models or models trained particularly on software artifacts (e.g., the work by \citet{panthaplackel-etal-2022-learning}).

\section{Understanding the Natural Practice of Summarization In Issue Threads}

In our initial exploration of long issue discussion threads, we observed cases in which participants provided summaries within the issues threads. In this section, we aim to understand this phenomenon more extensively and address our \textbf{RQ1: \textit{What are the characteristics of existing summaries that naturally occurred in the OSS issue discussion thread?}} We selected three OSS projects as a multiple case study to perform content analysis~\cite{Vaismoradi2013} on a sample of long issue threads to characterize the summaries and their roles within the issue discussions. Our analysis provides empirical evidence on the existing practice of summarization by issue users, as well as implications that served as a basis for the design of our tool.

\subsection{Methods}

\subsubsection{Data Selection}
For our analysis, we focused on the longest issue discussion threads sampled from three open source projects: \textit{Tensorflow}, a machine learning framework commonly used for deep learning applications, \textit{scikit-learn}, a general machine learning library, and \textit{Jupyter Notebook}, a web-based interactive development environment. We selected these three projects as case studies based on the following considerations. First, those projects are actively being developed and have a large community base. Projects at this scale are more likely to experience the problem of managing lengthy issue discussions and benefit from this work. Moreover, the domains of those projects align with the authors' technical knowledge; therefore, we can ensure the quality of the content analysis. From each selected project, we then sampled the ten longest issue threads. We focus our analysis on the longest issues because, again, participants from these threads are more likely to be overwhelmed by the discussion and, thus, more likely to adopt the summarization strategy in their comments.

\subsubsection{Identification of Summaries}
We adopted the following steps to identify the summaries included in the threads that we sampled. We define a \textit{\textbf{summary}} as \textit{a brief statement that expresses, in a concise form, information that appears elsewhere in the issue thread}.
First, we iteratively identified from our initial manual investigation a list of keywords that could indicate summaries (see Table~\ref{tab:summary_keywords}). We then constructed a regular expression with these keywords (including their variations, such as verb tenses, upper/lower cases, noun plurals, and spelling variations) and applied it to each issue discussion thread to identify sentences that contained the keywords; we considered those sentences as potential summaries. The keywords we used are not considered as an exhaustive list to capture all possible summaries. Instead, we tried to include all relevant words and phrases related to summaries identified during our manual investigation to increase the recall in the identification of true summaries.  
To understand the extent to which the keyword matching process can retrieve valid summaries, we performed a manual validation. Using six threads from the dataset, two from each repository, we manually identified all summaries, to create a list of gold standards. Compared with the gold standards, our regular expression displays a low precision of 17.02\% but a satisfactory recall of 88.89\%. This indicates that the keywords and the corresponding regular expression can retrieve most of the summaries, but requires further verification to ensure precision.

Next, we applied our regular expression to the remaining 24 issue threads and manually verified all the matched sentences by removing those that do not correspond to our definition of a summary. Through this process, we identified a total of 108 summaries in the 30 issue discussion threads.  

\begin{table}[ht]
\centering
\caption{Keywords we used for initial identification of summaries, categorized into four groups. These keywords are not considered to be an exhaustive list. Instead, they are iteratively identified to capture any relevant words or phrases from our manual investigation of summaries in issue discussions.}
\label{tab:summary_keywords}
\small
\begin{tabular}{p{5.5cm}p{7.5cm}}
    \toprule
    Category & Keywords\\
    \midrule
    Directly indicating summary & summary/summarize, sum up, outline, nutshell, all this to say, simply put, tl;dr\\
    Referring to a person or a comment & according to, other people, as others, suggested by, 's idea, we have, have tried, mention, talk, point, discuss,  comments, this issue, above, below, earlier, as indicated before, as I said\\
    Capturing a purpose related to summary & for future/reference/other, remember, recommend, reference, recap, recall, overview, updated/current list, list of remaining/done, conclude/conclusion, therefore/so, synthesize, that means\\
    Capturing an attribute or quality related to summary & short, quick, basically, high level, consensus, thus far, at this point, as far as, essence, exact, essential, underlying, biggest, important, relevant, overall\\
    \bottomrule
\end{tabular}
\end{table}

\subsubsection{Content Analysis}
Three authors of this paper performed three rounds of analyses to create a codebook and finalize the coding for all summaries in our dataset. Concretely, the authors first independently open-coded three randomly selected issue threads from the dataset (one from each project) to capture their initial observation of the summaries. They then discussed their codes together and created a preliminary codebook to capture the salient characteristics of the summaries. The codes were organized along three axes: (1) the \textit{objectives} of the summarization, (2) the concrete summarization \textit{style}, and (3) the \textit{target audience} of the summarization. During the second round, the three authors independently coded another three issues from the dataset using the preliminary codebook; then, they refined the codebook after discussing their disagreement on the coding. Finally, the three authors coded the entire dataset again using the updated codebook; the coding was conducted so that each issue thread was coded independently by two coders. 
We then used Cohen's Kappa~\cite{Cohen1960} to evaluate the agreement between the two coders. For the three axes of the codes (i.e., objective, style, and target audience), the Kappa values on categorical coding are 0.836, 0.844, and 0.838, respectively, indicating an ``almost perfect'' agreement~\cite{viera:garret:2005}. The three authors further discussed the cases in which there was a disagreement until they reached an agreement on all the summaries in the dataset. The codebook was further refined and finalized during this discussion.

\subsection{Results}
\label{sec:empirical_study_results}

\subsubsection{Overview}
We observed that manual summarization is common in most of the long issue threads: 28 out of 30 issues contain at least one summary. One thread (TensorFlow Issue \#22) even contains 13 summaries during its lifespan from November 2015 to June 2021. Sometimes, issue users would voice the necessity for providing the summary, such as ``\textit{It seems it's time for compressing the above discussion into one list again.}''\footnote{https://github.com/tensorflow/tensorflow/issues/22\#issuecomment-380402775} Other times, they phrase their frustration of \textit{not} having summarized information: ``\textit{Wait, I've been reading this thread/issue for 10 mins now. I got halfway through and I skipped through the rest. Are AMD GPUs supported yet?}''\footnote{https://github.com/tensorflow/tensorflow/issues/22\#issuecomment-323233294} Such frustration is not limited to the information seekers. It can be from the information providers or the project contributors moderating the thread: "\textit{It seems that you have not even bothered to read this thread, because there are several workarounds posted above including mine.}"\footnote{https://github.com/tensorflow/tensorflow/issues/22794\#issuecomment-445451227} 
While summaries can play an essential role in supporting the discussion in long threads, their characteristics can be context-dependent. Below, we discuss our observation of the primary categories of summaries along three axes, i.e., their objective, style, and target audience. The number of appearances of each category is indicated in the parenthesis following the name of that category.

\subsubsection{Objectives of Summarization}
\label{subsubsec:obej_sum}
Among all the analyzed summaries, we identified four categories of objectives for which a summary was created. 

\textbf{Add Context (32)}: Because the discussion on long issue threads can be entangled and even repetitive, commenters mention relevant content from the previous comments to supplement the current comment so that the readers would have the appropriate context. The commenters often use this kind of summary to report the outcome of their actions related to the software. For example, they tried a workaround or bug reproducing steps already mentioned in the previous comment and then reported in the new comment whether they obtained the same observation.

\textbf{Provide Access Point (31)}: The summary provides a starting point to make it easier for people to engage with the content. The commenters use this type of summary to highlight the most important information to the reader. Different from the category of \textit{Add Context} in which the summary is playing a supportive role to other new content, summaries in the category of \textit{Provide Access Point} normally are the main point of that paragraph or the entire comment. For example, the commenter often uses summary to recap the workaround of the issues under discussion: ``\textit{TL;DR -- just fully remove and reinstall Python with Tcl/Tk option selected}.''\footnote{https://github.com/tensorflow/tensorflow/issues/22794\#issuecomment-441431361}

\textbf{Provide Supporting Evidence (31)}: This type of summary backs up other actions, statements, or opinions within the same comment. The summary in this case acts as a reason or justification to explain ``why'' certain actions are taken, arguments are made, or opinions are voiced. For example, it was used when the maintainers closed the issue or provided judgment on the causes of the software bugs discussed in the issue. 

\textbf{Clarification (16)}: Depending on the topics under discussion, a single comment can already be dense and hard to digest. The commenters use this type of summary either to verify their own understanding of the previous comment by attempting to summarize concepts within the thread, or to make the concepts in their own comment easier for others to understand. They often use phrases such as ``\textit{to clarify what I meant}'' or ``\textit{is that right}'' before or after the summary to indicate the confirmatory nature of the summary.

\subsubsection{Style of Summarization} In terms of how the issue users write a summary, we observed four concrete styles. 

\textbf{Synthesize Facts (59)}: Instead of merely restating previous comments, the commenters summarize by synthesizing some content from the thread. While the summary may refer to one or more specific comments, the commenters build on the summary to draw new conclusions or extrapolate new insight. For example, in this comment, the commenter wrote ``\textit{This output seems unrelated to the discussion - which is focused on an issue relative to `pywin32'}.''\footnote{https://github.com/jupyter/notebook/issues/4909\#issuecomment-615894947} Here, the commenter quickly recapped the topic of this issue and judged the information provided in a previous comment as irrelevant. As the most commonly used style, this type of summary can be used to serve all four objectives listed in the previous section. 

\textbf{Remind and Recall (36)}: The commenters simply point to or restate content from previous comments without significantly altering the information contained. Similar to the style of \textit{Synthesize Facts}, this style is also used for various objectives. For example, in this case, the commenter indirectly quotes the content reported by others already to \textit{Add Context} when describing the nature of the bug: ``\textit{This is quite a weird bug. Like [User ID] mentioned, if I just reactivate the env by deactivating and activating again, it works fine, even though literally nothing else changed.}''\footnote{https://github.com/jupyter/notebook/issues/2359\#issuecomment-543169499} In another issue thread, the commenters quoted  the conclusion reached in earlier comments to \textit{Provide Supporting Evidence} for the action of locking the issue: ``\textit{To restate the conclusions for future readers: This is an issue with GitHub not with Jupyter notebooks.}''\footnote{https://github.com/jupyter/notebook/issues/3555\#issuecomment-517490294}

\textbf{Synthesize in Same Place (TL;DR) (7)}: The commenters provide a summary of their own content mostly to \textit{Provide Access Point}. The commenters normally provide a more detailed description of their own investigation or experience in the comment and use this type of summary before or after to isolate the key points. They often use phrases such as ``\textit{TL;DR}'' or ``\textit{long story short}'' to explicitly mark the summary.

\textbf{Set Up a Checklist (6)}. The commenters create a list of tasks to be accomplished which have not been discussed in the thread so far. The main purpose of this summary is to easily track the progress and focus on the remaining todos (i.e., \textit{Provide Access Point}). The commenters normally use different Markdown syntax to indicate the checklist, such as Strikethrough and Task List. Sometimes, they use an external file to keep the list and provide a link in the comment.\footnote{https://github.com/scikit-learn/scikit-learn/issues/3846\#issuecomment-315620842}

\subsubsection{Target Audience of Summarization}
Depending on the context of the summary, it might be written with a different audience in mind. We categorize the target audience into three groups.

\textbf{Participants in Discussion (62)}: This type of summary is meant to be read by more than one participant (other than the commenter) in the current thread. In most cases, commenters do not explicitly mention the name of the participants in the comment while sometimes they may tag the names of all the target participants. Since the role of the participants in each thread can be diverse, the summary might be intended exclusively for participants with a certain role, such as developers or software users depending on the information type of the summary (e.g., bug reproducibility and workaround). 

\textbf{A Specific Commenter (43)}: Very often, the issue comments serve as a conversational channel between a small number of participants on specific topics. The commenters often use the pronoun ``\textit{you}'' in their comment as an indication of response to another participant. To make the intent clear, they might further tag the name of the target participants explicitly in their summary. This type of summary normally responds to a comment posted close by in the thread if not immediately earlier.

\textbf{Future First-time Readers (3)}: Occasionally, the commenters provide the summary in particular for people who have not followed nor engaged in the discussion thread yet but would find the information useful. For example, the commenter for the Jupyter Notebook decided to close the issue because the resolution of the problem had already been reached in the previous comment but comments of similar requests were kept being posted by new users. In this case, the summary includes the target reader explicitly: ``\textit{To restate the conclusions for future readers: This is an issue with GitHub not with jupyter notebooks.}''\footnote{https://github.com/jupyter/notebook/issues/3555\#issuecomment-517490294}

\subsection{Discussion}
Manual summarization is a commonly-used strategy for users to participate  in discussions in long-living issue threads. As stated by some commenters in the issue threads, it might cause them great difficulty to appropriately digest the content in those threads without summaries. On the other hand, the explicit effort to provide such summaries can become under-appreciated given the current design of the issue thread. The summaries are simply buried within the overwhelming number of comments and fail to be caught by the intended readers. 

Moreover, the objective of summarization diversifies, ranging from supporting ongoing conversations with specific participants (e.g., \textit{Clarification}) to highlighting valuable information to current and future readers (e.g., \textit{Provide Access Point}). It is impossible to resort to any off-the-shelf summarization technique to satisfy all common use cases simultaneously. A summarization solution that tailors to the different needs of issue participants is necessary to support the OSS community. These observations motivate us to create a new design to facilitate the authoring and retrieval of summaries on issue threads that can serve various objectives while respecting the current usage of issue tracking systems in OSS development and maintenance.

\section{Design Process}
Our empirical case study illustrates the characteristics of the existing summary and how they are inadequately used because of the current ITS design. There is still a considerable gap between understanding the practice and proposing viable design options to support the summarization process and the best use of the resulting summaries. In this section, we fill this gap by answering \textbf{RQ2: \textit{What kind of support can be provided to satisfy the need to write and use summaries in OSS ITSs?}} Concretely, we lay down a set of design guidelines following a user-centered design process. First, we created the conceptual design and preliminary prototypes of a tool, named \toolname{}, based on our previous investigation and experience with issue discussion threads as well as existing literature. Then, we conducted a formative user study to evaluate the design concept and generate the design guidelines. We introduce how we follow these guidelines to finalize the design of our tool \toolname{} in Section~\ref{sec:design_impl}.

\subsection{Conceptual Design and Preliminary Prototyping}
The design concept of \toolname{} is informed by both our empirical investigation of existing issue threads and our own experiences and frustrations using the current ITSs. The overall goal of the tool is to \textit{facilitate collaborative authoring and usage of summaries to address the information seeking and acquisition challenges in issue discussions.} Thus, the tool should allow issue discussion participants to collaboratively write and edit summaries to satisfy their varied needs and objectives, as well as use the summaries to understand the often entangled issue conversations and efficiently retrieve relevant information. The tool should adopt means to support the summary authoring process, such as incorporating automated summarization techniques (e.g.~\cite{liu-lapata-2019-text}) to suggest summaries for user review, editing, and approval. The tool should also be easily integrated into the current issue discussion process to streamline its adoption and maximize its usage.

Particularly, we designed the tool for supporting two categories of summaries: (1) summaries that highlight a set of closely related comments (i.e., \textbf{conversation summaries}) and (2) summaries that encapsulate a certain aspect of the issue (i.e., \textbf{information type summaries}). The first category is inspired by the work of \citet{zhang2017wikum} about summarization in discussion forums. The summaries sit alongside the issue thread to provide the most important information, with the original context immediately available for cross-reference. The necessity of this type of summary is also supported by the objectives of \textit{Add Context} and \textit{Clarification} observed in our empirical investigation. For the second category, we are inspired by the work of \citet{Arya2019}, which revealed that open source issue discussions are often embedded with various common types of information. We adopted the classification schema of information types identified in their work (e.g., information related to expected behavior, motivation, solution discussion, workarounds, etc.) and used the automated information type detection model they provided. This type of summary can serve the objectives of \textit{Provide Access Point} and \textit{Provide Supporting Evidence}, both are common for the existing manual summaries (see Section~\ref{subsubsec:obej_sum}).

Based on these design concepts, we created a preliminary set of sketches and UI prototypes of the tool. We designed the tool to supplement the current GitHub Issues UI so that OSS participants can easily plug it into their current workflow. We envision that the tool can be adapted to other ITSs with reasonable effort. In our preliminary design, the tool will be loaded once the user opens an issue thread. The \textit{conversation summaries} are located next to the issue discussion thread as additional comments expandable to reveal the original comments that they summarize. The \textit{information type summaries} are fixed to the top of the page as access points, before the issue report and all comments. Each information type that is detected in the issue discussion has a separate space for displaying its summary. ITS users would be able to trigger the generation of both types of summaries, as well as review and edit them. The users would also be able to manually adjust the automatically detected information types in the issue thread. These sketches and prototypes are used in the formative user study to further assess user needs and distill concrete system design guidelines.

\subsection{Formative User Study}
\subsubsection{Methods}

We conducted a formative study with three target users to understand how they digest information within issue discussion threads, their frustrations with current systems, and their design preferences for supportive tools. The study is approved by the research ethics boards of all involved universities. The participants were recruited from personal networks and social media ads. The three participants varied in their frequency of use of GitHub Issues and their type of contribution to the open source projects, representing a diversity of backgrounds. 

\begin{itemize}[noitemsep,topsep=0.5pt]
    \item P1 is self-identified as an expert user of GitHub Issues, who frequently reads issues and has created or commented on issues at least once per month. They use the ITS mostly when they experience problems with an open source library or tool and try to find solutions or seek help from the development teams. They also provide code contributions to open source projects occasionally.

    \item P2 is a casual user of GitHub Issues, who reads issues sometimes but has never contributed to the discussion. They use the ITS mostly when they experience problems using open source projects and want to find confirmation or workarounds. They have never provided code contributions to any project. This participant represents typical OSS end-users.

    \item P3 is self-identified as a code contributor and a code reviewer. They use the ITS mainly to track the discussion around their development work. They frequently open and comment on issues, but tend to disregard issues irrelevant to their work at hand.
\end{itemize}

Each study session lasted about 60 minutes, for which the participants were compensated with \$15 CAD. During the sessions, we first asked the participants to describe their current usage of ITSs, the challenges that they face in their usage, and their experiences in encountering and writing summaries in issue discussions. Then, we presented our preliminary prototype of the \toolname{} tool and asked the participants to articulate (1) general feedback on the tool's concept, (2) the perceived usefulness of the conversation summaries and the information type summaries in the participants' current workflow, and (3) reasons for using or not using this tool when available; the tool prototype is used here as a probe to help elicit the needs of the participants. Once finished, the user study sessions were fully transcribed and analyzed using an inductive thematic analysis approach~\cite{Vaismoradi2013}. We discuss the salient themes of their challenges, needs, and feedback for a summarization tool in the following sections.

\subsubsection{Current Experience with the Issue Discussion Threads}
The participants shared several common challenges when engaging in issue discussions and adopted different strategies to overcome these challenges.

\textbf{\textit{Understanding discussion threads is difficult.}} Across the range of their backgrounds, all participants have faced difficulty comprehending issue discussion threads. Some participants mentioned that it is \textbf{hard to follow the rationale} why certain actions and directions in the conversation were taken. As P2 put it, ``\textit{You have to sort of try and judge what was done, what was ignored here, what was completely changed, and that's when ... confusion starts to come up}.''
Another issue arises when the participants in the discussion \textbf{misunderstand the focus} of the thread. For example, P1 mentioned a potential problem that ``\textit{people talk about something not necessarily helpful in the thread.}'' 
Threads are also made more confusing by \textbf{entangled conversations}. For example, P3 mentioned, ``\textit{I have seen that so many people are talking about different things in the comments, and sometimes people ... are commenting other stuff.}'' Participants also expressed frustration with GitHub Issues'  \textbf{lack of support for obtaining information} within issues. For example, P1 mentioned, ``\textit{[Searching] is really hard because you need to do control F ... then scroll down to see which match.}'' Combined together, these challenges reflected the fact that certain comments or information are buried in the issue threads and not made easily obtainable for the users. 

\textbf{\textit{Different strategies are adopted to find information in issue threads.}}
All participants expressed that they do not read long issue discussion threads in their entirety, opting for alternative ways of finding the portions of the thread that are relevant to them. The most straightforward strategy for looking for information in those threads is to \textbf{skim the thread} until they find the relevant portion. For example, when looking for a specific topic, P2 mentioned their approach: ``\textit{It's mostly just scrolling down into where that was being discussed. There are not that many ways to filter out posts.}'' Participants sometimes \textbf{relied upon social signals} to find relevant posts. For example, P3 said, ``\textit{If there are too many [comments] to read, I just check the comments which have the most ... thumbs up.}'' This could indicate that getting an overview of the thread quickly is key to users of ITSs. Participants particularly found that \textbf{summaries, although rare, are helpful} in this process. For example, P1 mentioned that ``\textit{For the thread that's long, people try to keep track of what's going on ... I don't need to scroll all the way up to see what people have been doing.}'' This confirmed that discussion participants were already making efforts to make a uniform account of the thread to keep everyone up to speed, and that users may be receptive to this.
P2 also emphasized the benefits of summaries, saying, ``\textit{It makes you avoid having to read through everything -- if you have a summary of what they're saying. It could also tell you whether or not it's worth reading ... all of their conversations.}''

\subsubsection{Needs and Feedback of a Summarization Tool}
With the help of the initial prototype, participants voiced various needs for a summarization tool for facilitating the usage of ITSs and provided feedback on our design concept. 

\textbf{\textit{Depending on the usage scenario, users find different types of summaries useful.}}
Some participants suggested that the information type summaries are helpful for users when they have a particular search goal. For example, P1 said ``\textit{I think the top feature [information type summaries] is very useful ... because we never have that -- you need to do Ctrl-F and it is not even matching the comment, not searching word by word like Google.}'' Some participants considered the information type summaries as an overview of the discussion thread. As P3 puts it, ``\textit{I personally prefer one summary to summarize all of the stuff. I'm a lazy person. I just want to know what's in the issue -- is there any solution, and then if there's any suggestion. That's it.}'' Since they are only interested in certain categories of information, the information overview is an efficient way of getting a big picture of the thread. At the same time, participants mentioned that using the information type summaries can risk losing the contexts and the details. For example, P2 mentioned: ``\textit{I sometimes have a very peculiar setup on my computer ... you need to understand what's going on, and you can't understand if you just blindly follow a series of commands that was posted on the issues page.}'' Many times, the participants mentioned that their unique situation shapes their perceived usefulness of different types of summaries. As P2 discussed, ``\textit{The conversation level summaries ([conversation summaries]) seem a lot more specific. ... However, [the information type summaries] give you a good general idea of what's going on.}'' Therefore, we conclude that the diversity of user goals is best supported by having both types of summaries.

\textbf{\textit{Need to minimize required effort.}}
Participants expressed a preference for a design that does not require too much manual effort and does not interfere with their normal use of GitHub Issues. For example, in our prototype, users were prompted to manually confirm the information types detected in their comment before posting it.
When asked about this feature, P3 said ``\textit{It's useful to the application, but I'm lazy. I don't want to do that ... There's a lot of steps. Even if I ask to automatically generate, I still need to check if the words are correct or not and then select the corresponding comments.}.'' 
Participants also suggested lowering the amount of effort required to contribute to summaries. For example, P2 observed that people often do not make the effort to organize information on GitHub Issues. They said automatically generated summaries would be useful ``\textit{because the manual ones sort of depend on people actually creating them and maintaining them, which isn't always the case, and which they might not always do. And so I think the automated ones would be more useful to have since it's lower maintenance and it's sort of always going to update itself.}''

\textbf{\textit{Motivation to contribute is a critical consideration.}}
Participants expressed the importance of motivating users to contribute to the summaries. This is related to the previous point that there is a concern about the human effort and collaboration required for summarization. For example, P3 had the following question: ``\textit{I'm wondering how to make more people to collaborate and contribute their time to labeling this stuff.}'' They identified the recognition of contributions as central to this question of motivation, suggesting ``\textit{GitHub has activity blocks showing you how much code you write today or how many pull requests you had today. This summarization could be counted as their contribution as well, visually showing on their page, and so people may be more interested in doing that.}'' P1 also suggested using social features to motivate people to contribute, saying ``\textit{For example, similar to Facebook ... people can come in as a top fan. Then people will think ah, she's following everything. They will have an impact.}'' These suggestions center around public recognition of individual participants in the discussion and summarization to motivate contributions .

\subsection{Design Guidelines}
We distilled our investigation of summarization from the empirical study of existing summaries and formative user studies into a set of design guidelines. These guidelines, summarized in Table~\ref{tab:design_guidelines}, are created to inform the design of tools that leverage collaborative summarization to support information seeking and acquisition in OSS issue discussions. 

\begin{table}[ht]
\centering
\caption{Design guidelines distilled from the content analysis of natural summaries and the formative user study.}
\label{tab:design_guidelines}
\small
\resizebox{\textwidth}{!}{
\begin{tabular}{p{4cm}p{5.5cm}p{4.5cm}}
    \toprule
    \textbf{Design Guideline} & \textbf{Evidence$^{*}$} & \textbf{Potential Design Actions} \\
    \midrule
    
    GL1: The tool should be flexible to accommodate various objectives and use cases of summarization &
    \begin{minipage}[t]{\linewidth}
    \begin{itemize}[leftmargin=*]
        \item[$\blacktriangleright$] Various types of objectives for summarization were identified
        \item[$\vartriangleright$] Participants described different scenarios of using summaries
    \end{itemize}
    \end{minipage}&
    \begin{minipage}[t]{\linewidth}
    \begin{itemize}[leftmargin=*]
        \item Provide different types of summaries
        \item Allow users to toggle different summarization views
    \end{itemize}
    \end{minipage}\\
    
    \\
    GL2: The discussion context need to be preserved in the summaries. &
    \begin{minipage}[t]{\linewidth}
    \begin{itemize}[leftmargin=*]
        \item[$\blacktriangleright$] \textit{Add Context} and \textit{Provide Supporting Evidence} are common summarization objectives.
        \item[$\vartriangleright$] Issue discussions often include entangled conversations.
        \item[$\vartriangleright$] Participants find following the rationale of a certain argument challenging.
    \end{itemize}
    \end{minipage} &
    \begin{minipage}[t]{\linewidth}
    \begin{itemize}[leftmargin=*]
        \item Preserve the chronological thread view of comments.
        \item Link the summaries to the comments that they summarize.
    \end{itemize}
    \end{minipage}\\
    
    \\
    GL3: Important information within the thread should be made readily available. &
    \begin{minipage}[t]{\linewidth}
    \begin{itemize}[leftmargin=*]
        \item[$\blacktriangleright$] \textit{Provide Access Point} is a common summarization objective.
        \item[$\blacktriangleright$] \textit{Synthesize Facts} is a common style of summarization.
        \item[$\vartriangleright$] Participants mentioned challenges to understand the focus and search for useful information.
    \end{itemize}
    \end{minipage} &
    \begin{minipage}[t]{\linewidth}
    \begin{itemize}[leftmargin=*]
        \item Summaries need to capture the diverse aspects in the issue discussion.
        \item Summaries need to be easily visible.
    \end{itemize}
    \end{minipage}\\
    
    \\
    GL4: The system should support iterative management of summaries. &
    \begin{minipage}[t]{\linewidth}
    \begin{itemize}[leftmargin=*]
        \item[$\blacktriangleright$] Natural summaries appeared throughout the discussion thread, at different discussion stages.
        \item[$\vartriangleright$] Summaries can be used in many usage scenarios.
    \end{itemize}
    \end{minipage} &
    \begin{minipage}[t]{\linewidth}
    \begin{itemize}[leftmargin=*]
        \item Allow users to create summaries at any point of discussion.
        \item Summaries should be always modifiable.
    \end{itemize}
    \end{minipage}\\
    
    \\
    GL5: Users should be encouraged to contribute to collective sense-making efforts. &
    \begin{minipage}[t]{\linewidth}
    \begin{itemize}[leftmargin=*]
        \item[$\vartriangleright$] Natural summaries in issue threads are useful but rarely available.
        \item[$\vartriangleright$] Minimizing manual effort and increasing motivation to contribute are important user needs.
    \end{itemize}
    \end{minipage} &
    \begin{minipage}[t]{\linewidth}
    \begin{itemize}[leftmargin=*]
        \item Recognize contributions to creating and editing summaries.
        \item Use automated techniques to reduce workload.
    \end{itemize}
    \end{minipage}\\ \\
\bottomrule
\end{tabular}
}
{\raggedright $^{*}$ \footnotesize
 $\blacktriangleright$ \textit{Evidence from the content analysis of natural summaries } \newline
\hspace*{1.5mm} $\vartriangleright$ \textit{Evidence from the formative user study}
 \par}
\end{table}

\section{System Design and Implementation}
\label{sec:design_impl}
We developed a working version of the \toolname{} tool based on the prototypes following the previously outlined design guidelines. As an extension for Chromium-based browsers, the tool allows seamless integration into a user's workflow when accessing the ITS. In the following sections, we describe the interaction design and the implementation of the tool. The design guidelines that contributed to the design are indicated in brackets.

\subsection{Overall System Design}
Enabling the extension in the browser shows the two additional UI components for \textit{Conversation Summary} and \textit{Information Type Summary} to the page [GL1]; see Figure~\ref{fig:summit_overview}. The \textit{information type summary} panel is visible in an expanded view at the top of the thread, and changes its form to appear as a navigation bar and sticks to the top of the viewport while the user scrolls through the thread [GL3]. The \textit{conversation summary} panel is visible on the left side and sticks to the side while scrolling [GL3]. It is possible to increase the size of \textit{conversation summary} if needed [GL3]. When an issue thread is introduced to the tool for the first time, both \textit{conversation summary} and \textit{information type summary} are blank; the summaries are populated on request.

\begin{figure}[t]
    \centering
    \includegraphics[width=0.9\textwidth]{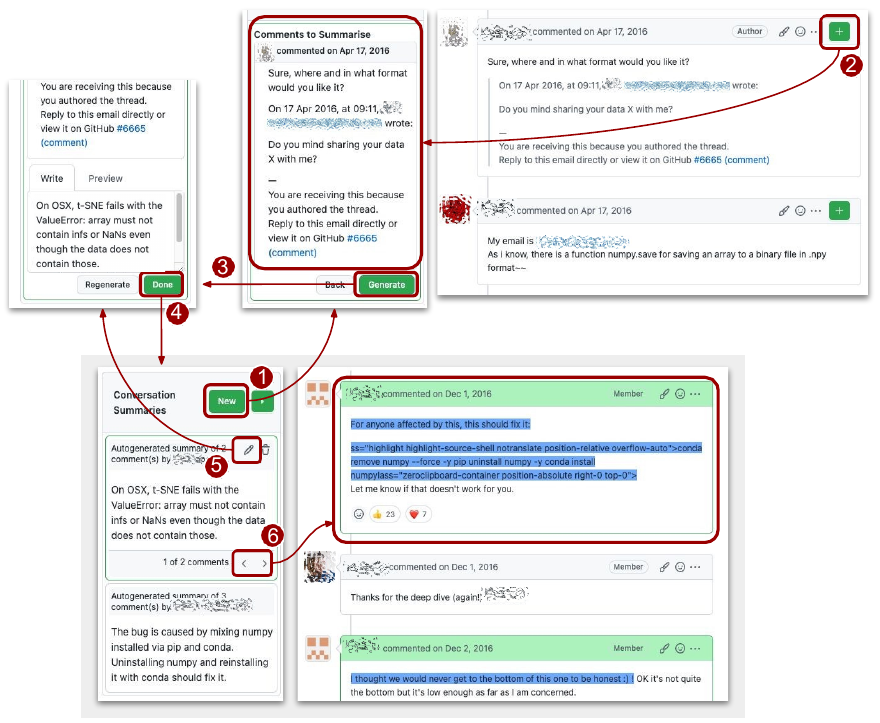}
    \caption{User interaction flow of the conversation summary component.}
    \label{fig:conv_summary_flow}
\end{figure}

\subsection{Conversation Summary}
The \textit{conversation summary} allows the users to summarize a set of comments of their choosing; Fig.~\ref{fig:conv_summary_flow} illustrates the user interaction flow of this component. The panel defaults to a \textit{summary} view and displays an instruction about creating the \textit{conversation summaries} if there are no summaries present. Clicking on the ``New'' button (\circled{1} in Fig.~\ref{fig:conv_summary_flow}) in the \textit{conversation summary} component adds a button on every issue comment indicating that it can be added to a summary (\circled{2} in Fig.~\ref{fig:conv_summary_flow}). The button is greyed out if a summary is already created using that particular comment. Once the first comment is added, the panel switches to \textit{comment} view that shows all the comments to be summarized [GL2]. If a comment is no longer desired in the summary, then it can be discarded directly from the panel, or by clicking on the ``X'' button on the issue comment. The panel also contains a ``Generate'' button which on click sends a request to the backend with the added comments (\circled{3} in Fig.~\ref{fig:conv_summary_flow}).

Clicking the ``Generate'' button switches the panel to a \textit{edit summary} view that contains the comments, a text box element with the summary response in markdown format from the backend and a preview panel that renders the contents of the text box. The summary content can be modified if needed and then saved [GL5]. An option to regenerate the summary is provided [GL5]. Once the summary is saved  (\circled{4} in Fig.~\ref{fig:conv_summary_flow}), the panel goes back to the default \textit{summary} view and displays all the summaries while mentioning the authors of the comments which each is summarizing. If there are multiple summaries for an issue, they are displayed as an unordered list [GL3]. 

On clicking a summary, the issue comments that make up a \textit{conversation summary} are highlighted and the thread auto-scrolls to the first issue comment that contributed to the summary [GL2]. Several other interactions are provided, such as an edit button that switches the panel to the \textit{edit summary} view (\circled{5} in Fig.~\ref{fig:conv_summary_flow}) [GL4], a delete button that deletes the summary [GL4], and navigation buttons for browsing through the issue comments (\circled{6} in Fig.~\ref{fig:conv_summary_flow}) [GL2]. 

\begin{figure}[t]
    \centering
    \includegraphics[width=\textwidth]{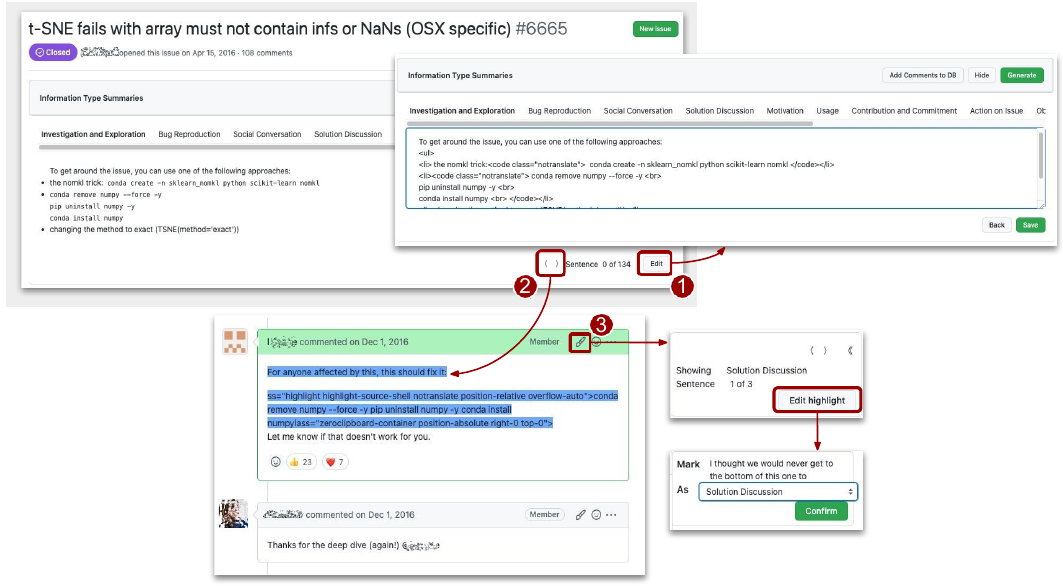}
    \caption{User interaction flow of the information type summary component.}
    \label{fig:info_summary_flow}
\end{figure}

\subsection{Information Type Summary}
The \textit{information type summary} panel displays summaries of information types~\cite{Arya2019} in a tab view format; the major user interaction flow of this component is illustrated in Fig.~\ref{fig:info_summary_flow}. In order to generate the summaries on request, a button is provided to send a background request to add the issue comments to the database. The first instance of the background request adds the existing comments to the database and subsequent requests update the database with any new comments added to the thread. A ``Generate'' button is provided to generate (or regenerate) the summaries from the existing comments in the database. 

The predicted information type summaries are shown in separate horizontal tabs at the top of the issue thread; the tabs are sorted in descending order according to the number of sentences in each information type [GL3]. The model generated summaries for each information type can be edited by clicking on the ``Edit'' button provided in each tab (\circled{1} in Fig.~\ref{fig:info_summary_flow}). Clicking on the ``Edit'' button provides a text box that supports markdown formatting, a ``Back'' button to discard changes, and a ``Save'' button to save them [GL4]. Once an edit is made, the summary authors' names are shown at the bottom of the summary [GL5]. 

Each tab provides navigation buttons to navigate to the corresponding sentences in the issue thread that contribute to the information type summary (\circled{2} in Fig.~\ref{fig:info_summary_flow}) [GL2]. When the thread is scrolled, the summary panel minimizes into a navigation bar that sticks on top of the thread for easy access. The bar contains the navigation buttons, a drop down menu to choose the information type to navigate, and shows the number of sentences in the chosen information type. The sentences of the specific information type are highlighted to distinguish them from the rest of the issue thread [GL2]. An edit button is provided on top of every issue comment that allows uses to manually mark a sentence as belonging to an information type if it is unmarked or incorrectly classified as a different information type by the model (\circled{3} in Fig.~\ref{fig:info_summary_flow}). A ``Hide'' button is provided to minimize the panel if the user does not prefer to use \textit{information type summaries} [GL1]. 

\subsection{Implementation}
\label{subsec:implementation}
The tool comprises a browser extension developed with the React framework (https://reactjs.org) and a backend system developed with Python and FastAPI (https://fastapi.tiangolo.com). The extension adds the SUMMIT-specific UI elements to the GitHub issue discussion web page. These elements were developed with Primer (https://primer.style), GitHub's design system, in order to conform with the existing GitHub UI theme and components. 

The backend consists of two services -- an \textit{extension handler} and a \textit{summarization engine}. The extension handler is hosted on the Heroku platform (https://www.heroku.com) and leverages the Postgres database (https://www.postgresql.org) for storing the issue comments and associated summaries. The main purpose of the handler is to serve the necessary CRUD (create, read, update, and delete) functionality supported by the extension, detecting sentence boundaries, and predicting the information type of the sentences. The sentences in any incoming request are preprocessed by tagging and removing the hyperlinks, code blocks, and markdown syntax labels before sentencizing them using the SpaCy sentencizer module (https://spacy.io/api/sentencizer). The information type of each sentence is inferred using the model described in ~\citet{Arya2019}. The \textit{summarization engine} is a Google Cloud Platform CloudFunction (https://cloud.google.com/functions) that is invoked on an HTTP request from the \textit{extension handler}. The engine consists of a specialized preprocessor that tags and removes hyperlinks, code blocks, and markdown syntax labels from the sentences to be summarized. The sentences cleaned in this way are then fed to the BERTSumExt model~\cite{liu-lapata-2019-text} which produces an extractive summary. Once the summary is generated, the previously removed contents (i.e., hyperlinks, code blocks, and markdown labels) are restored into the sentences extracted in the summary. Then the summary is sent as a response to the \textit{extension handler}. 

The \textit{extension handler} relays the summary back to the extension which displays it as an information type summary or as a prompt in the case of a conversation summary. If the request is for information type summaries, calculated sentence boundaries are returned and are used by the information type summary component to highlight the sentences in the thread. Since the summaries are maintained in a centralized database, anyone who has installed the extension and navigates to the issue thread is able to view them.

\section{User Evaluation}
In this Section, we answer our \textbf{RQ3: \textit{How does \toolname{} affect the OSS stakeholders' behavior and perception of the information acquisition process?}} Since issues are used by different OSS stakeholders in various contexts, we conducted a summative lab user study to observe our participants in two common use cases on GitHub with and without \toolname{}. This approach allows us to study the participants in a controlled environment that is close to real-world practice~\cite{PiggybackPrototyping}. In particular, we investigate the following three concrete aspects during the user study:

\begin{itemize}
    \item \textbf{RQ3.1}: How does the issue users' behavior differ when \toolname{} is presented compared with when it is not?
    \item \textbf{RQ3.2}: How would using \toolname{} affect the users' perceived task difficulty and their confidence when completing common tasks on issue threads?
    \item \textbf{RQ3.3}: What are the users' perceived usefulness and associated risks of \toolname{}?
\end{itemize}

\label{sec:summative_user_study}
\subsection{Methods}
Below, we describe our participants, the study procedure, and the analysis methods. The study was approved by the research ethics boards of all involved universities.

\subsubsection{Participants}
We aim to recruit participants with experience both in open source software development and the domain of the software so that their behavior during the study can resemble the information acquisition process in practice. Therefore, we selected one project from our case studies (i.e., scikit-learn) and performed targeted recruitment. We first posted ads on online groups and forums about open source and machine learning, the social media channel of the authors' universities, and the personal networks of the authors. While we listed the required background in the ads, we further ensured the fitness of the participants through a pre-study survey to confirm that they have used GitHub Issues as one of the following roles: user, code contributor, code reviewer, designer, and project maintainer/owner. We also asked if they had previous data science experience since the issues used in the user studies were related to a machine learning library, scikit-learn.  A total of 16 participants joined our study, referred to as P2-P17\footnote{P1 refers to the participant who joined our pilot user study and the outcome was excluded from the final analysis.}. Among them, seven were female and nine were male. Participants self-identified to have various roles in their recent involvements in OSS communities, including six participants identified as project maintainers/owners, four as code contributors, four as code reviewers, and two as designers. All participants were familiar with GitHub Issues and they read, reported, or commented on issues at least twice a week. All participants also self-reported having data science experience to understand technical aspects of related issue discussions.

\subsubsection{Procedure}
Each user study session took about 60 minutes to complete, for which the participants received a compensation of \$15 CAD. We first introduced the basic features of \toolname{} to the participants and then asked them to install and freely interact with it. Afterwards, we asked the participants to perform two similar sets of tasks on two issue threads from the scikit-learn project\footnote{https://github.com/scikit-learn/scikit-learn/issues/6665 and https://github.com/scikit-learn/scikit-learn/issues/2889}, with the tool support on one thread and without on the other. Those two issue threads were chosen because they are both long threads with complex information exchanges among issue commenters. The summaries presented on the tool were auto-generated by SUMMIT in advance with minor manual edits on coherence and conciseness. The order of these two conditions (i.e., using the tool vs. not using the tool) and the condition-issue assignments were counterbalanced using Zeelenberg and Pecher’s approach~\cite{Zeelenberg2015}. 

The two tasks that the participants were asked to perform on each thread represent two common use cases for software users and contributors, respectively, when interacting with issue discussion threads. \textbf{The first task} is to find a workaround within the issue thread. The participants were asked to act as users of scikit-learn who were required to use an older version of the library and were experiencing the problem described in the issue post. The participants, therefore, needed to find a workaround that does not involve upgrading the version of the software. \textbf{The second task} is to understand the existing investigation of the bug. The participants were asked to determine if the cause of the bug is related to certain factors (i.e., the Python version in one issue and the API parameters in the other). Prior to the study, none of our participants have read the issues selected. Therefore, their behavior on the first task of each thread also represented what a first-time issue reader might exhibit. When proceeding to the second task on the same issue, they might have already developed a certain understanding of the relevant topics, therefore can represent a returning reader of the thread. In this study, we ask the participants to perform those tasks in the provided environment (with and without \toolname{} being presented) without instructing them \textit{how} they should perform those tasks (in particular which part of the tool to use when \toolname{} is presented). This is a deliberate decision to closely mimic the realistic information acquisition process and allow us to observe how the different parts of the tool might support those use cases in practice.

Each task was capped for ten minutes to manage the total effort of the study; if the participants could not provide a decisive answer by the end of this period, we asked them to give their best guess. To ensure the privacy of the participants, they were asked not to make public comments or other actions on GitHub when performing the tasks. After each task, the participants rated on a five-point Likert scale about how challenging they considered the task and their confidence in their answers. They were also asked to provide a brief explanation for each of the above ratings. 
Once all tasks were completed, we conducted a short exit interview with the participants to understand their experience when completing the tasks and their feedback on the tool.

\subsubsection{Analysis}
For the participants' ratings, we conducted statistical analysis to compare the two conditions (i.e., using \toolname{} vs. not using \toolname{}) in each task.  
Since the data are essentially ordinal and we cannot assume a normal distribution of the measures, we used the Mann-Whitney U test.

Additionally, we performed an inductive thematic analysis~\cite{Vaismoradi2013} on the qualitative data, which includes the process they performed the tasks with and without the tool support as well as their feedback during the post-study interview. 
Particularly, two of the authors first inductively coded two random videos individually. They then discussed and agreed on a coding strategy. The remaining videos were then divided in half; each independently coded by one of the authors. They then discussed and merged their coding to identify common themes.

\subsection{RQ3.1 Results: Information Acquisition Behavior}
Between 75\% and 87.5\% of the participants obtained the accurate answer in the tasks; we did not find a statistically significant difference in the task success status when the tool was presented versus not. When performing the tasks, however, our participants adopted a diverse set of strategies to obtain the relevant information depending on whether our tool was presented. We describe these strategies below.

\subsubsection{Information Acquisition Strategies with \toolname{}}
We categorized the participants' strategies for acquiring information with the support of \toolname{} along the degree to which they relied on the summaries versus the actual comments. We found four levels of reliance from observing how participants performed the tasks. 

First, when summaries were available, some participants \textbf{relied only on the information presented in the summaries} to perform the tasks (P2, P5, P8, P11, P12, P17 in Task 1 and P2, P3, P11, P14, P15 in Task 2). These participants focused only on the summaries, either the information type summaries or the conversation summaries, or both
and did not read the comments other than the original issue post. Second, some participants \textbf{used the summaries with the support of the linked comments/sentences} to verify the summaries, understand the context, and/or retrieve richer information (P3, P4, P6, P7, P9, P13, P14 in Task 1 and P8, P9, P12 in Task 2). They usually started by reading the summaries. When finding an interesting summary, they used the navigation features provided by the tool to inspect the related comments or sentences. Sometimes, they also read comments around the highlighted sentences/comments to understand the context and verify their understanding. Third, some participants did not use the summaries at all and \textbf{relied solely on the comments even though the summaries are present} (P10, P15, P16 in Task 1 and P6, P7, P10, P16 in Task 2). While they did not directly interact with the summaries, they sometimes used the highlights related to a certain information type or a certain conversation topic provided by our tool to navigate the discussion thread and direct their attention. Finally, when performing Task 2, some participants \textbf{relied on their memory or previous knowledge} and did not read the comments or the summaries (P4, P5, P13, P17).

When the participants used the summaries (i.e., in the first two cases above), some only focused on one type of summaries. In Task 1, five participants (P2, P11, P12, P13, P14) only used the information type summaries, while three (P5, P6, P9) only used the conversation summaries; five others (P3, P4, P7, P8, P17) used both types of summaries. In Task 2, two participants (P12, P15) used only the information type summaries, while four only used the conversation summaries (P2, P3, P9, P14) and two used both (P8, P11). 

\subsubsection{Information Acquisition Strategies without \toolname{}}
Participants attempted various strategies to obtain the information when the summaries were not presented, which we grouped into the following categories; some participants adopted multiple strategies. First, a common strategy was to \textbf{read through the comments chronologically}. The majority of participants ($N=14$) used this strategy for Task 1. For Task 2, while about half of the participants ($N=7$) used this strategy, four of them (P4, P10, P12, P16) scrolled through the comments quickly and only focused on the interesting and relevant ones they remembered from Task 1. Second, some participants (P3, P5 in Task 1 and P3, P9, P16 in Task 2) \textbf{used the browser's searching function} to find keywords in the hope to obtain relevant information. Interestingly, some participants (P5, P7, P14, P16, P17 in Task 1 and P16 in Task 2) \textbf{used various discussion signals} to narrow down to the relevant comments. These strategies included: (1) focusing on comments with more emoji reactions from the community members, (2) focusing on the most recent comments, and (3) following comments made by a certain author such as the issue poster or an author who made useful comments before; occasionally, these comments included the natural summaries written by the commenters in the original issue thread. Finally, in Task 2 about half of the participants ($N=7$) \textbf{relied on memory or previous knowledge} to complete the task; this number is larger than when the tool is used.

\subsection{RQ3.2 Results: Perceived Task Difficulties and Confidence}
We found that the participants rated Task 1 as significantly less challenging when using the tool ($median=2$), compared to not using the tool ($median=4$), confirmed by a Mann-Whitney U test ($U(16,16)=39$, $p<0.001$). When explaining their challenge ratings, participants mentioned several rationales that we grouped into two main categories. First, the participants found that \textbf{using the tool reduced the number of comments to examine}. For example, P5 mentioned ``\textit{[With the tool,] I can have some focus. And by navigating the main points I could do my own research in those details from comments through here. So it's not that a stressful to go over all the comments.}'' P7 also commented that ``\textit{the tool is useful to highlight the important story.}'' Second, participants mentioned that \textbf{the target information was sometimes readily available in the summaries}. For example, P2 mentioned that Task 1 is ``\textit{as easy as going to one of the tabs in the information type summary panel.}'' P17 also attributed their rating to ``\textit{the fact that I was able to get the solution from the plugin directly rather than scrolling.}'' On the contrary, participants usually found \textbf{frustrating to understand the complexity of the issue without the tool support}. Many participants mentioned that there were too many comments to read and understand. For example, P9 said, ``\textit{I couldn't tell really which one would help because there are a bunch of different solutions to it. It's hard to find it manually.}'' Participants also considered the issue thread as convoluted, as P6 put: ``\textit{I didn't realize that some of them were not really related.}'' P4 also mentioned: ``\textit{Following along human conversation with back and forth with lots of [comments] before the problem has been identified -- that's like most of the early comments -- is not useful.}'' Interestingly, although P14 used the social signals to quickly find the most promising comments, they still rated the task as neutral and considered this approach as ``\textit{just guessing, not verified}.''

On the other hand, the challenge ratings on Task 2 were generally lower than Task 1 ($median=2$ with the tool and $median=2.5$ without the tool); the difference between the two conditions for Task 2 is not statistically significant ($U(16,16)=84$, $p<0.101$). Many participants attributed the decreased challenge level in Task 2 to the fact that they got familiar with the issue after performing the first task, as P2 put ``\textit{This [Task 2] is a lot easier than the last one [Task 1], since I have already familiar with the issue.}'' However, the participants still considered that the process was generally more smooth with the tool support; for example, P3 commented: ``\textit{[With the tool,] the answer is right in front of me here [as a conversation summary]. It is very accessible. This is like one of those Google snippets that when you Google something, you can get the answer quickly.}''

Additionally, participants were fairly confident that they completed both tasks successfully, regardless of using the tool or not ($median=4$ for both tasks in both conditions). Participants usually considered that the confidence originated from the concrete evidence they found in the comments, as P9 put, ``\textit{The comments I found pretty succinctly say that they figured out [the cause].}'' Sometimes, the comments resonated with each other, which increased confidence; for example, P3 was confident about their answer because ``\textit{a lot of comments said that -- a lot of people verified it.}'' On the contrary, when participants were not confident about their answers, the reasons given were often related to the fact that they were not able to verify their solution themselves. For example, P10 was unsure about all their answers because ``\textit{This is written by other people and I did not test the solution yet.}'' P12 was also not confident about an answer because ``\textit{the solution may not be universal.}''

\subsection{RQ3.3 Results: Perceived Usefulness and Risks of \toolname{}}
\subsubsection{Overall Experience}
Comparing performing the tasks with and without \toolname{}, the participants discussed several key differences they perceived in the following aspects.

\textbf{Mental effort:} Going through a long issue discussion thread is an extremely demanding task. P7 mentioned that without the tool, they need to budget a lot of time and mental effort to go over the task. P9 further attributed the mental effort to the gap between the large volume of information presented in the thread and the capacity of human attention: ``\textit{[without the tool] it is almost intimidating because it was very manual and I felt like I would past what I was trying to find half the time and it was very annoying.}''  With the tool, on the other hand, ``\textit{it is definitely more manageable}'' (P7). 

\textbf{Information Organization:} Different from the linear way of accessing the information presented in the thread, the tool offers a global view of the discussion in the thread. Such restructuring of information greatly impacts how the participants get oriented in the issue thread. As P17 suggested: ``\textit{I found myself to have had more useful information right up front when I opened the thread as compared to [without the tool]}". Similarly, P2 said ``\textit{without the tool it took a lot longer to look through all the comments because you don't know which ones will be verified by posts from other people. Whereas with the tool it automatically checks all the posts for relations and what they said and auto generates, which makes it a lot easier to find the solution.}''

\subsubsection{Information Type Summaries Versus Conversation Summaries} Our participants expressed different opinions on the usefulness of the two types of summaries. Some of our participants considered both types of summaries useful, although they might be useful in different situations. For example, P3 described two specific use cases for each type of summary: ``\textit{if I were just to get a glimpse of the document before I dive in and try to find what I'm looking for, then the conversation summaries are great for that, and if I'm trying to narrow down the scope of my search, then the tabs [of information type] are great for that.}'' P9 suggested that the information type tabs would be useful in the long run because it includes more diverse information while the conversation summary helped them more for the tasks during the study. 

Most of the other participants have a clear preference for one type of summaries over the other. Information Type summary is considered the most useful feature by P10, P11, and P17. For example, P17 suggested that the information type summary is the first place they would check: ``\textit{I would look at the top [information type summary], perform all the things that I can see on top. If not, I'll jump to the sidebar [conversation summary] to see if there's anything that users mentioned that seems to be useful. If both of them failed, then my third fallback would be to read the thread as it was without the plugin.}'' The reason for such preference is that the information type tabs allow them to \textbf{prioritize their effort} on the more important content such as solution discussion (both for P10 and P17).  P11 stated ``\textit{more exhausted [coverage] of the issue}'' is another reason they consider the information type summary more useful.

On the other hand, P2, P5, P6, P7, and P13 considered the conversation summary more useful. The reason is mostly attributed to the fact that the target of the summary is normally a short conversation within the issue thread. The information is not as compressed as the information type summary is and, therefore, the participants can easily \textbf{verify the summary quality} supported by the comment navigation function of \toolname{}. 

\subsubsection{Risks}
When being asked about their trust in the summaries provided by \toolname{} and the potential risks associated with adopting the tool in their workflow, our participants have taken views from the following aspects.

\textbf{Summary Quality:}
While most of the participants considered incorrect/unuseful content in the summary as one of the biggest risks, their opinion diverged on the severity of this risk. P11 considered it as the deciding factor about adopting the tool: ``\textit{it should be 100\% accurate or I will not come back to the tool. I would rather read the entire thread rather than get something wrong. [I] would put more trust on the summary than the comment so if it is not true, it would be a deal-breaker.}'' On the contrary, P7 thought this is not a big issue considering the content from original posts can contain problems, to begin with. In general, most participants stated that they would manually verify the content found from the issue thread by themselves regardless of whether it is from the summary or the original issue thread.

\textbf{Impact on User Behavior:} 
Two participants (P4 and P7) voiced the potential risk of changing the behavior of programmers when they overly rely on the tool. They considered the complexity of the issue thread as an essential and useful training material. The summaries might hide the complexity and deprive the programmers of the opportunities to improve their skills in navigating multi-facet issues. They believe the tool should be primarily used as signposts for the full context of the issues.

\textbf{Model Performance:} Since SUMMIT uses the machine-learning based natural language processing models to perform the information type classification and summarization (see Section~\ref{subsec:implementation}), participants raised their concerns about slow inference or incorrect model output that might negatively impact the user experience. Furthermore, if humans fail to correct the generated summary, it would lead to the risks mentioned above related to \textit{Summary Quality}.

\section{Discussion}
Overall, our study delineated a comprehensive picture of the role of summarization in issue tracking systems (ITSs) for open source software (OSS). Through an empirical investigation of natural summaries in existing issue threads and a formative user study with different OSS stakeholders, we distilled a series of design guidelines for tools aimed at helping OSS stakeholders get oriented in issue threads anchored in summaries. These guidelines were subsequently used in the creation of the \toolname{} tool. The findings of our summative user study revealed that using \toolname{} has (1) drastically reshaped the information acquisition strategies of the participants in long issue threads, (2) helped surface useful content from the thread and decreased the perceived difficulties in finding relevant information, and (3) reduced the mental effort of comprehending long issue threads while improving information organization. Together, our findings demonstrated the potential of \toolname{} and the corresponding design guidelines in supporting users to acquire information from lengthy discussion threads in ITSs. Below, we reflect on our results through the key guidelines that we originally proposed and discuss the implications of our findings for designing similar tools.

\subsection{Reflections on GL1: Be flexible to accommodate different goals and preferences}
Our empirical investigation of natural summaries in the existing issue threads and our formative user study uncovered the multifaceted nature of summarization in the ITS, indicating the importance of providing flexible support for different summarization goals and preferences. These findings were further echoed in our summative user studies. When completing the tasks, participants differed in their preference using either or both types of summaries (i.e., \textit{conversation summaries} and \textit{information type summaries}) depending on their roles in the OSS community and their tasks at hand; some participants only used the highlight function that the \toolname{} tool supported even when summaries are available. In their feedback, participants clearly indicated the importance and benefits of having both types of summaries. Together, our results demonstrated the value of creating flexible collaborative tools for communities with diverse backgrounds such as OSS communities. The two types of summaries included in \toolname{} provide the foundation for building summarization tools for such communities.

\subsection{Reflections on GL2: Preserve discussion context}
In our empirical study and formative user study, we found that summaries are usually used for providing relevant context and emphasizing important evidence to support arguments or decisions within an entangled issue thread, indicating the importance of preserving discussion context in summaries. When designing \toolname{}, this guideline is incorporated into the feature that provides easy navigation from the summary to the specific comments captured by the summary. In our summative user study, we found that this feature is frequently used by the participants to find discussion contexts. Participants also suggested that the comments corresponding to the summary are more likely to be important for issue readers compared with other content in the thread. Moreover, they frequently mentioned that finding the corresponding content within context (i.e., the comments linked to the summaries) is an important source of confidence. Overall, our results indicated that preserving discussion context in summaries supported users in gaining valuable contextual information and judging the credibility of the summaries on collaboration platforms such as the ITS. Such content verification activity is important especially when the exchange is abundant, distributed, and generated by various humans and machines, such as the summaries in our case.

\subsection{Reflections on GL3: Make important information readily available}
Summarization, from a general point of view, is to extract and synthesize important information from a large amount of content. We have identified the need of making such information readily available to ITS users in the empirical study and the formative user study. The \toolname{} tool is thus designed to meet the need by making such information always visible on the screen when users read through the issues threads. This design has facilitated our participants to quickly find relevant information from lengthy threads. In fact, a considerable number of participants only used the information presented in the summaries to complete the tasks. Participants also frequently mentioned that they can effortlessly find the answers from the summaries, which made the tasks easier to complete with \toolname{}. Additionally, since the summaries, in particular, the information type summaries, cover various important aspects of the discussion, making them always available also provides a comprehensive overview of the issue threads, which is appreciated by our participants. However, an intricate question arises concerning the ``importance'' of information. Particularly, participants discussed the risk of losing the perspective related to the complexity of an issue when using summarization tools. In other words, using summarization blindly may make users overlook some nuanced information represented in the discussion thread. Considering such a risk, we believe that this guideline should be better integrated with the other guidelines we proposed to facilitate flexible judgement, iterative negotiation, and collective sense-making regarding the importance of information in the discussion context.

\subsection{Reflections on GL4: Support iterative management of summaries}
This guideline originally encompassed our findings that natural summaries in issue threads appeared at different points of discussion and may be used in various scenarios. We thus designed \toolname{} to allow users freely create and update summaries at any point in the lifetime of an issue thread. It also serves the purpose of content moderation to avoid errors made by users or the underlying summarization model. Many participants also highlighted the need for content moderation and expressed their hesitation to trust the summary quality and model performance. The discussion context (captured in our GL2), therefore, was considered critical for the participants to perform sanity checks and to inspect the context of the summary in detail when necessary. Thus, when applying this guideline in reality, especially with automated tools, it is crucial to support users to correct and update the content. Additionally, considering GL3, a mechanism for granting privileges of summary creation and modification should be carefully examined. Future work is needed to explore these points.

\subsection{Reflections on GL5: Encourage collective sense-making}
Issue discussion is innately a collaborative activity. Facilitating and encouraging diverse community members to engage in this collaborative endeavor is key for collective sense-making. In our summative user study, we frequently observed participants using social signals (e.g., \textit{``thumb up''} and \textit{``+1''} emoji reactions left on a comment) to judge the quality and relevance of a comment. Such signals, however, are not always available. Moreover, the efforts sometimes go unrecognized because of how the existing ITSs are designed. \toolname{} offers a new way for the OSS community members to make more noticeable contributions. Some of our participants voiced their willingness to add and improve the summaries and suggested that those efforts would benefit the project in the long term. Extending this point of view, considering community inclusiveness in tool design is a critical factor to truly encourage collaborative sense-making~\cite{costanza2020design}. In our case, GitHub is primarily tailored to developers and project managers with related technical backgrounds. Other OSS stakeholders such as end-users and designers often have great difficulty navigating various features, including the issue discussions~\cite{jaz_chi_ea}. During our empirical study, we observed many occasions where these marginalized stakeholders expressed struggle and frustration when trying to find relevant information and make meaningful comments. Correspondingly, the automated summarization function in \toolname{} can provide a draft version of the summary, and therefore, lower the contribution barrier. On the other hand, related to our reflection on GL3 and GL4, there is a delicate balance between motivating/enabling the contribution and controlling the quality of the contribution. While the flexibility and auto-summarization features of \toolname{} help alleviate the contribution effort, it might also permit errors when users are not careful when using the tool. Thus, future work should explore ways to both incorporate the needs of the marginalized stakeholders and ensure contribution quality in tools like \toolname{}.

\section{Limitations and Future Work}
Our study has the following major limitations. First, in the empirical investigation of the current summarization practice, we only focused on long threads from three OSS projects. This focus is considering that long issues are more likely to overwhelm discussion participants and, therefore, more likely to contain summaries. The three OSS projects are selected based on the fact that they have large and diverse community bases. However, the generalizability of our results needs to be verified in future work.

Second, the sample size of our formative user study that partially contributed to the design guidelines is relatively small. Because of the small sample size, we selected participants who have varied experiences and profiles with respect to the use of GitHub Issues. This satisfied the exploratory purpose to identify the main challenges and needs of users in order to triangulate with the findings from analyzing natural summarization practices, which in turn resulted in the design of \toolname{}. In the future, conducting a larger-sample user study may expose other user challenges and needs that can be satisfied by tool design.

Third, our tool is designed as a plug-in on top of the GitHub issue tracking system. This decision was made to maximize our potential impact within a manageable scope since GitHub is one of the largest social coding platforms. However, the applicability of our approach in other open source discussion platforms needs to be further explored.

Fourth, our tool currently uses an automated summarization technique that relies only on texts presented in the discussion comments when generating summaries. Metadata associated with an issue, such as issue resolution and closure, are not automatically captured in the summaries. However, in our design, all automatically generated summaries can be edited by users. So, if relevant information in the metadata is important, users can decide to manually include it in one or more summaries. To extend our work, future studies can explore techniques to incorporate both textual data and metadata in summaries.

Finally, the summative user study was conducted in a lab setting, which limits the complexity of the tasks. The tasks we used may also be different from the ones that the participants encounter in their own projects. The users' behavior also might deviate from using the tool in real-world settings. However, the lab study format allows an experimental design that gives us more control over a consistent structure to observe users' behaviors and collect their feedback. In our future work, we plan to gather evidence on the long-term impact of using collaborative summarization tools like \toolname{} to support OSS issue discussions.

\section{Conclusion}
Making sense of issue threads and finding useful information from them is a demanding task for users of open source software (OSS) issue tracker systems (ITSs). Our empirical investigation of long issues from popular OSS projects indicated that summarization is an effective way to accelerate the information acquisition process and serves multiple objectives in ITSs. However, summaries take considerable effort to write and can be overlooked by the issue users due to the design of current ITSs. We proposed a set of design guidelines describing concerns that a summarization tool for ITSs should address as well as several concrete design options. Through an iterative user-centered design process, we developed \toolname{} following the design guidelines. \toolname{} supplements the existing ITSs with two types of summaries that re-organize the information from the issue thread: \textit{conversation summaries} highlight the key points of a set of related issue comments and \textit{information type summaries} provide an index-like structure to issue threads. The user studies with 16 participants demonstrated the impact of \toolname{} on tasks commonly performed using OSS issues. \toolname{} lowered the mental effort of the issue users and allowed them to prioritize their attention. Our work illustrates the potential of building crowd-based and machine-learning-enabled tools to support the summary authoring and information acquisition process on ITSs.

\begin{acks}
We thank the participants of our user study for their valuable time and thoughtful feedback. We also thank the anonymous reviewers for helping us improve the paper. This work is partially supported by the Alfred P. Sloan Foundation (Grant No.: G-2021-16745).
\end{acks}

\bibliographystyle{ACM-Reference-Format}
\bibliography{reference}

\end{document}